\documentclass[aps,prl,noshowpacs,superscriptaddress,amsmath,amssymb,twocolumn,preprintnumbers,reprint]{revtex4}
\usepackage[utf8]{inputenc}
\usepackage{graphicx}
\usepackage{xcolor}
\usepackage{bm}
\usepackage{hyperref}
\hypersetup{colorlinks=true,linkbordercolor=blue,linkcolor=blue, citecolor=blue,pdfborderstyle={/S/U/W 1}}
\usepackage{xspace}

\def\be{\begin{equation}}
\def\ee{\end{equation}}
\def\bea{\begin{eqnarray}}
\def\eea{\end{eqnarray}}
\newcommand{\ave}[1]{\langle{#1}\rangle}
\newcommand{\eq}[1]{(\ref{#1})}
\newcommand{\jpsi}{\ensuremath{J/\psi}\xspace}
\newcommand{\D}{\ensuremath{D}\xspace}
\newcommand{\dd}{{\rm d}}
\newcommand{\sqrts}{\sqrt{s}}
\newcommand{\pt}{\ensuremath{p_{_\perp}}}
\newcommand{\dsigpp}{\dd\sigma_{\rm pp}}

\newcommand{\RAA}{\ensuremath{R_{{\rm AA}}}}

\newcommand{\qhat}{\hat{q}}
\newcommand\meanz{\langle z \rangle}
\newcommand\meanepsbar{\ensuremath{\bar{\epsilon}}\xspace} 
\newcommand{\meaneps}{\ensuremath{\langle\epsilon\rangle}\xspace}
\newcommand{\TeV}{\ensuremath{\,\text{Te\hspace{-.08em}V}}\xspace}
\newcommand{\GeV}{\ensuremath{\,\text{Ge\hspace{-.08em}V}}\xspace}
\newcommand{\dNch}{\ensuremath{\frac{\dd N_{\rm ch}}{\dd y}}\xspace}
\newcommand{\dNg}{\ensuremath{\frac{\dd N_{k}}{\dd y}}\xspace}
\newcommand{\alphas}{\ensuremath{\alpha_s}\xspace}
\newcommand{\AT}{\ensuremath{A_{\perp}}\xspace}
\newcommand{\qzero}{\ensuremath{\hat{q}_0}\xspace}
\newcommand{\PbPb}{\ensuremath{\text{PbPb}}\xspace}
\newcommand{\XeXe}{\ensuremath{\text{XeXe}}\xspace}
\newcommand{\AuAu}{\ensuremath{\text{AuAu}}\xspace}
\newcommand{\ecc}{\ensuremath{\textnormal{e}}\xspace}

\begin{document}
\title{Probing the path-length dependence of parton energy loss \\[0.2cm] via scaling properties in heavy ion collisions}
\author{Fran\c{c}ois Arleo}
\affiliation{SUBATECH UMR 6457 (IMT Atlantique, Universit\'e de Nantes, IN2P3/CNRS), 4 rue Alfred Kastler, 44307 Nantes, France}
\author{Guillaume Falmagne}
\affiliation{SUBATECH UMR 6457 (IMT Atlantique, Universit\'e de Nantes, IN2P3/CNRS), 4 rue Alfred Kastler, 44307 Nantes, France}
\affiliation{Laboratoire Leprince-Ringuet, CNRS/IN2P3, \'Ecole polytechnique, Institut Polytechnique de Paris, Palaiseau, France}
\affiliation{High Meadows Environmental Institute, Guyot Hall, Princeton University, Princeton, NJ 08544-1003, USA}

\date{\today}
\begin{abstract}
The scaling property of large-$p_\perp$ hadron suppression, $R_{\rm{AA}}(p_\perp)$, measured in heavy ion collisions at RHIC and LHC leads to the determination of the average parton energy loss $\langle \epsilon \rangle$ in quark-gluon plasma produced in a variety of collision systems and centrality classes. Relating $\langle \epsilon \rangle$ to the particle multiplicity and collision geometry allows for probing the dependence of parton energy loss on the path-length $L$. We find that $\langle \epsilon \rangle \propto L^\beta$ with $\beta=1.02{\raisebox{0.3ex}{\scriptsize$\substack{+0.09\\-0.06}$}}$, consistent with the pQCD expectation of parton energy loss in a longitudinally expanding quark-gluon plasma. We then demonstrate that the azimuthal anisotropy coefficient divided by the collision eccentricity, $v_2/\textnormal{e}$, follows the same scaling property as $R_{\rm{AA}}$. This scaling is observed in data, which are reproduced by the model at large $p_\perp$. Finally, a linear relationship between $v_2/\textnormal{e}$ and the logarithmic derivative of $R_{\rm{AA}}$ is found and confirmed in data, offering an additional way to probe the $L$ dependence of parton energy loss using coming measurements from LHC Run 3. 
\end{abstract}
\maketitle

\setcounter{footnote}{0}
\renewcommand{\thefootnote}{\arabic{footnote}}

The theory of parton energy loss in quark-gluon plasma (QGP) and its associate jet quenching phenomenology in heavy ion collisions have become increasingly mature over the past decade, triggered by RHIC and LHC measurements with unprecedented precision and variety. While in the 2000s the attention had been put on the quenching of single hadron spectra and di-hadron correlations, the focus has then naturally shifted towards jet observables, as a result of new experimental measurements and theoretical ideas (see Refs.~\cite{Cao:2020wlm,Cunqueiro:2021wls} for recent reviews). Although related in principle, high-$\pt$ hadrons and jets are likely to probe different aspects of medium-induced gluon radiation. The former appears as a good proxy of genuine \emph{parton} energy loss in QGP, as originally designed in the theoretical formalisms~\cite{Baier:1996kr,Baier:1996sk,Gyulassy:1999zd,Gyulassy:2000er,Wiedemann:2000za,Arnold:2000dr}; the latter, instead, probes the gluon emission off a final state made of multiple particles acting coherently~\cite{Blaizot:2015lma}. Despite these advances, fundamental questions remain. Among them, how parton energy loss in QGP depends parametrically on the medium path-length $L$~--~addressed in many studies~\cite{Shuryak:2001me,Bass:2008rv,Dominguez:2008vd,Chesler:2008uy,PHENIX:2010nlr,Betz:2014cza,Noronha-Hostler:2016eow,Djordjevic:2018ita}~--~is discussed here.

We pursue in this Letter the approach initiated in Ref.~\cite{Arleo:2017ntr}, aiming at the understanding of large-$\pt$ hadron production in heavy ion collisions within a data-driven strategy based on a simple analytic energy loss model. Despite obvious limitations, this philosophy may reveal physical properties in data such as scaling laws. This was the case in Ref.~\cite{Arleo:2017ntr} where $\RAA$ is shown to be a universal function of $\pt/\meanepsbar$, with $\meanepsbar$ being the average parton energy loss \meaneps in a given collision system, multiplied by the mean fragmentation variable, $\meanepsbar = \meanz \meaneps$. Starting from this result, confirmed presently using additional data sets, we first explore the relation between $\meaneps$ and the multiplicity density $n_0$, $\meaneps\propto n_0\,L^\beta$, eventually allowing us to extract the parametric $L$-dependence of parton energy loss in QGP. We then use this dependence to determine the azimuthal anisotropy coefficient divided by the collision eccentricity, $v_2/\ecc$. This ratio follows the same scaling property as $\RAA$, as confirmed in data. Finally, a simple relation between $v_2/\ecc$ and $\RAA$ is found, offering a novel and data-driven way to probe the path-length dependence of parton energy loss.

Let us start with the analytic energy loss model. The nuclear modification factor at large $\pt$ is a scaling function of $\pt/\meanepsbar$,~\cite{Arleo:2017ntr}
\be\label{eq:RAA2}
\RAA^{h}(\pt, \meanepsbar, n)=f(u\equiv \pt/\meanepsbar, n)\,,
\ee
with $f$ given by
\bea\label{eq:f}
f(u, n)&=&
 \int_0^{\infty}\dd{x}\  \bar{P}(x)\,\left( 1 + \frac{x}{u} \right)^{-n}\\
 &\simeq& \int_0^{\infty}\dd{x}\  \bar{P}(x)\,\exp \left( - \frac{n x}{u} \right)\label{eq:u/n_scaling}\,,
\eea
where $n=n^h(\sqrt{s})$ is the spectral index of the pp production cross section, $\dsigpp^h/\dd\pt \propto \pt^{-n}$. The rescaled quenching weight, $\bar{P}(x=\epsilon/\meanepsbar) \equiv \meanepsbar\,P(\epsilon)$, is computed in~\cite{Baier:2001yt,Arleo:2002kh} from the BDMPS medium-induced gluon spectrum~\cite{Baier:1996kr,Baier:1996sk}. The scaling behavior~\eq{eq:RAA2} can be observed at a given collision energy, while the approximate scaling in  $\pt/n \meanepsbar$, Eq.~\eqref{eq:u/n_scaling}, allows for comparing $\RAA$ at different energies and for different hadron species. 
\begin{table}[t]
{\footnotesize
 \centering
 	\def\arraystretch{1.4}
 \begin{tabular}[c]{cccc}
\hline
\hline
Species & Collision & $\sqrt{s}$~(\TeV) & Experiment\\
$\pi^0$ & \AuAu & $0.2$ &  PHENIX~\cite{PHENIX:2012jha}\\
$h^\pm$ & \PbPb & $2.76$ & ALICE~\cite{ALICE:2018vuu}, ATLAS~\cite{ATLAS:2015qmb}, CMS~\cite{CMS:2012aa} \\
$h^\pm$ &  \PbPb & $5.02$ & ALICE~\cite{ALICE:2018vuu}, CMS~\cite{CMS:2016xef}\\
$h^\pm$ &  \XeXe & $5.44$ & CMS~\cite{CMS:2018yyx}\\
\hline
$D$ & \PbPb & $ 5.02$ & ALICE~\cite{ALICE:2018lyv}, CMS~\cite{CMS:2017qjw}\\
\hline
$\jpsi$ & \PbPb & $5.02$ & ATLAS~\cite{ATLAS:2018hqe}, CMS~\cite{CMS:2017uuv}\\
\hline
\hline
\end{tabular}
 \caption{$\RAA$ data used in this Letter.}
 \label{tab:data}
 }
\end{table}

\begin{figure}[t]
\begin{center}
    \includegraphics[width=0.47\textwidth]{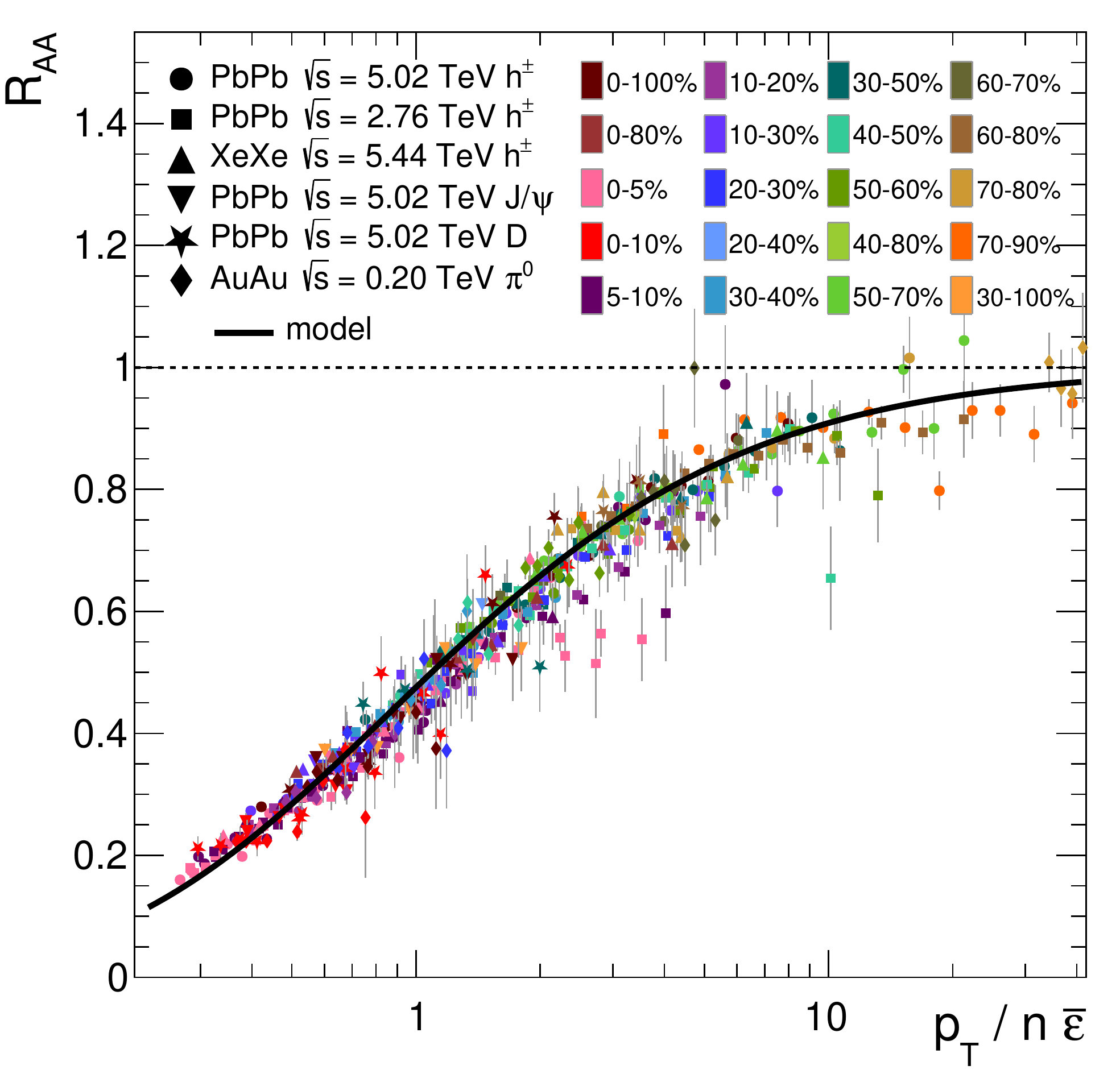}
  \end{center}
\vspace{-0.4cm}
\caption{Scaling of $\RAA$ of light hadrons ($h^\pm$ and $\pi^0$), $D$ and $\jpsi$ as a function of $\pt/\,n\meanepsbar$, in various collision systems.}
   \label{fig:scaling}
\end{figure}

In this model, the {\it shape} of $\RAA$ as a function of $\pt$ is thus fully predicted once $n$ is obtained from a fit to pp data. With respect to Ref.~\cite{Arleo:2017ntr} more systems (\XeXe at $\sqrts=5.44\TeV$, \AuAu at $\sqrts=200\GeV$) and  additional $\RAA$ measurements of $h^\pm$, $\jpsi$ and \D mesons have been included. The selection and geometric bias affecting $\RAA$ in a given centrality class is taken into account through a correction factor~\cite{Loizides:2017sqq}, which does not exceed 4\% for centralities below 50\%. All the measurements (listed in Table~\ref{tab:data}) with a $\pt\gtrsim10\GeV$ cut are plotted in Fig.~\ref{fig:scaling} as a function of $\pt/\,n\meanepsbar$~\footnote{For clarity, only statistical uncertainties are shown and data points with uncertainties larger than 0.1 are removed.}, where $\meanepsbar$ is fitted from $\RAA$ in each collision system (\textit{i.e.} nuclei species, collision energy, and centrality class). Clearly, $\RAA$ data line up as predicted into a single universal curve consistent with the shape of $\RAA$ given by~\eq{eq:RAA2}, shown as a solid line. This is consistent with a unique process being responsible for the nuclear modification factors of all hadrons above a given $\pt$.

Our goal is now to relate $\meanepsbar$ to the relevant physical quantities in heavy ion collisions. In the BDMPS formalism, the average parton energy loss in QGP can be written as~\cite{Baier:1996kr,Baier:1996sk}
\be
\label{eq:mean_eloss_expanding}
    \meaneps = \frac{1}{4}\,\alphas\, C_k\, \ave{\qhat} \,L^2\,,
\ee 
where $L$ is the average medium path-length and $C_k$ is the color charge of the parton ($C_q=4/3$, $C_g=3$). The gluon transport coefficient $\ave{\qhat}$ is linearly averaged along the parton trajectory,~\cite{Salgado:2002cd}
\be
	\ave{\qhat} = \frac{2}{L^2} \int_{\tau_0}^{\tau_0+L} \dd \tau \, (\tau-\tau_0) \, \qhat(\tau) \,,
\ee
taking into account the dynamical expansion of the medium produced at time $\tau_0$. 
The transport coefficient being proportional to the decreasing medium parton density, its time evolution can be parametrized as $\qhat(\tau) = \qhat_0 \left(\tau_0/\tau\right)^{\alpha}$, leading to
\be\label{eq:qhatalphaDep}
\ave{\qhat} = \frac{2}{2-\alpha} \, \qzero\, \left(\frac{\tau_0}{L}\right)^{\alpha}
\ee  
for $L \gg \tau_0$. The initial transport coefficient \mbox{$\qzero = \qhat(\tau_0)$} and parton density $n_0$ are directly related, \mbox{$\qzero = (9\pi/2)\,\alphas^2\,n_0$}~\cite{Gyulassy:2000gk}. In the Bjorken picture, $n_0$ can be estimated as~\cite{Bjorken:1982qr}
\be\label{eq:densitydNch}
n_0 = \frac{1}{\AT \tau_0}\,\left.\dNg\right|_{y=0} \,=\,\frac{3}{2}\,\frac{1}{\AT \tau_0}\,\left.\dNch\right|_{y=0} \, ,
\ee
where $\AT$ is the transverse overlap area of the two crossing nuclei. The rightmost equality assumes local parton-hadron duality ($N_k = N_h$) and the factor $N_h/N_{\textnormal{ch}}=3/2$ takes into account that a third of the produced particles (mostly pions) are electrically neutral.
Putting all together leads to~\cite{Gyulassy:2000gk}
\be\label{eq:eloss_scaling}
\meanepsbar = K\,\times\left(\frac{1}{\AT}\,\dNch \,L^{\beta}\right) \,,
\ee
with $\beta=2-\alpha$ and $K=27\pi/(8\beta)\times\alphas^3\,\tau_0^{1-\beta}\,\ave{z}_{_k}\,C_k$.
%

In the following, we check that the scaling relation Eq.~\eqref{eq:eloss_scaling} indeed holds for light hadrons in all collision systems. The geometric quantities $A_\perp$ and $L$ entering Eq.~\eqref{eq:eloss_scaling} are determined through an optical Glauber model, assuming hard sphere nuclear densities. The average path-length in the transverse plane is given by~\cite{Loizides:2017ack}
\bea\label{eq:L}
L &\equiv& 2 \int \dd\bm{\ell}\,\dd\bm{x}\, 
\rho_\textnormal{coll}(\bm{x})
\,\rho_\textnormal{part}(\bm{x}+\bm{\ell})\, |\bm{\ell}|\nonumber\\
&&\,  \Big/\, \int \dd\bm{\ell}\,\dd\bm{x}\, 
\rho_\textnormal{coll}(\bm{x})
\,\rho_\textnormal{part}(\bm{x}+\bm{\ell})\,,
\eea
where $\rho_\textnormal{part}$ and $\rho_\textnormal{coll}$ are the transverse distributions in the number of participants and of binary nucleon-nucleon collisions, respectively~--~the latter being the distribution of hard parton production points. The charged particle multiplicity $\dd N_\textnormal{ch}/\dd \eta$ at mid-rapidity is taken from PHENIX measurements~\cite{PHENIX:2015tbb} at RHIC and from \mbox{ALICE}~\cite{ALICE:2015juo,ALICE:2010mlf,ALICE:2018cpu} and CMS~\cite{CMS:2011aqh,CMS:2019gzk} measurements at LHC, multiplied by $J=d\eta/dy=1.25$ at RHIC~\cite{PHENIX:2004vdg} and $J=1.09$ at LHC~\cite{CMS:2010tjh}.
\begin{figure}[t]
\begin{center}
    \hspace*{-2mm}\includegraphics[width=0.47\textwidth]{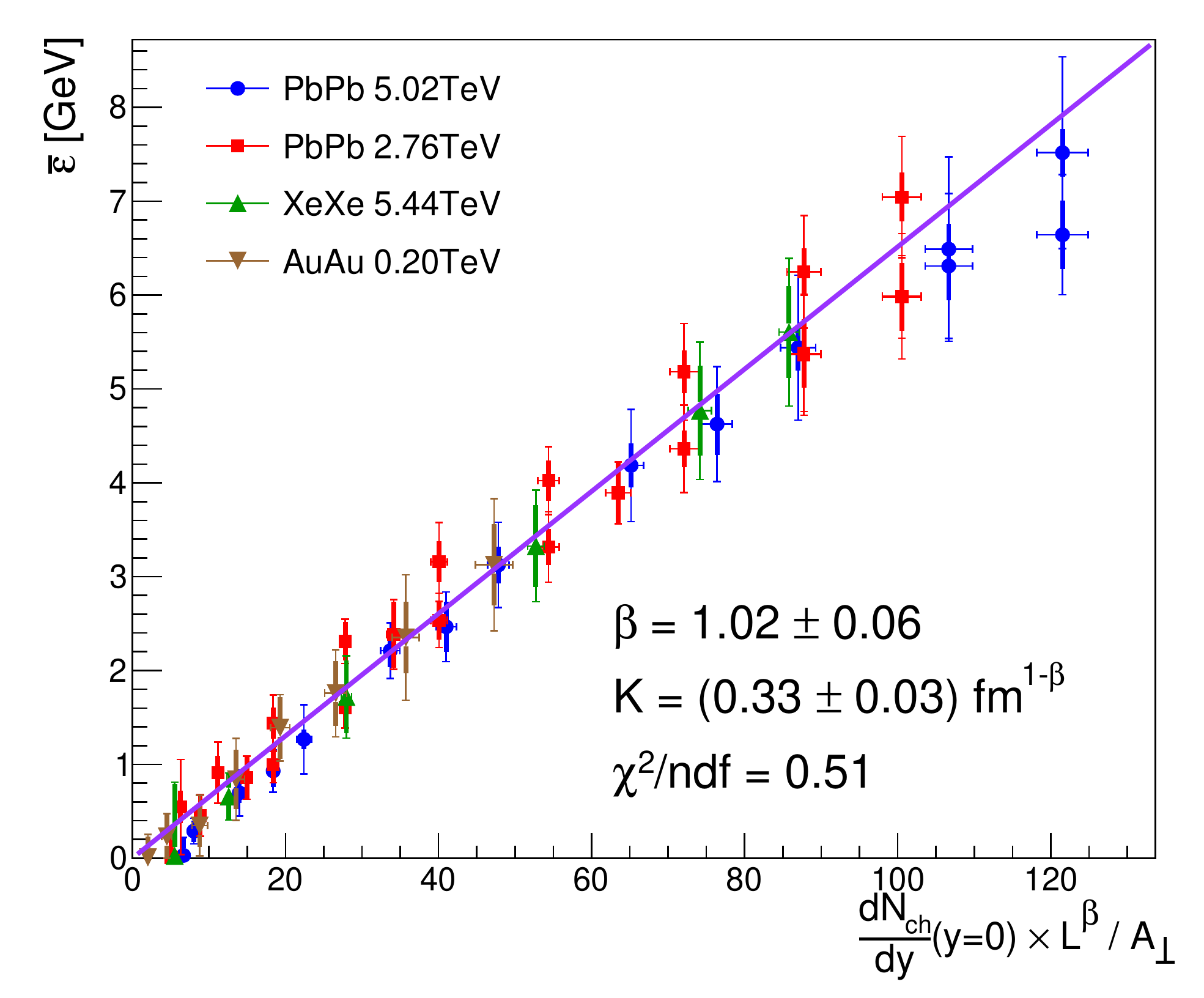}
  \end{center}
\vspace{-0.4cm}
\caption{Average parton energy loss extracted from $\RAA$ of light hadrons in various collision systems as a function of $\dd N_\textnormal{ch}/\dd y\times L^{\beta}/\AT$.}
   \label{fig:scaling2}
\end{figure}

The energy loss scales $\meanepsbar$ extracted from the quenching of light hadrons are fitted using Eq.~\eqref{eq:eloss_scaling}, with $K$ and $\beta$ taken as free parameters. Fig.~\ref{fig:scaling2} exhibits the excellent agreement ($\chi^2/\textnormal{ndf}=0.51$) obtained with a linear dependence of $\meanepsbar$ with the scaling variable $\dd N_\textnormal{ch}/\dd y\times L^{\beta}/\AT$. The fits leads to $K=0.33{\raisebox{0.5ex}{\tiny$\substack{+0.11\\-0.03}$}}$~fm$^{1-\beta}$ and $\beta=1.02{\raisebox{0.5ex}{\tiny$\substack{+0.09\\-0.06}$}}$. The uncertainties originate from the fit and from the use of alternative Glauber models for the calculation of $L$ and $\AT$: using constant $\rho_\textnormal{part}$, Woods-Saxons nuclear densities, or taken from a Glauber Monte Carlo model~\cite{Loizides:2017ack}. 

The value of $\beta$ proves compatible with unity, that is the BDMPS expectation in QGP experiencing a purely longitudinal expansion (i.e. $\alpha=1$). In particular, it seems to exclude $\meaneps \propto L^{3-\alpha}$ expected at strong coupling~\cite{Dominguez:2008vd,Chesler:2008uy}, at least with reasonable values of $\alpha$. Turning to the parameter $K$, its theoretical expectation depends on several uncertain quantities, e.g. the value of $\alphas$ ($K\propto \alphas^3$), the nature of the propagating parton (hence its color charge~\footnote{A linearly-averaged color charge is expected when the fragmentation of both quarks and gluons is considered.}), and the fragmentation variable. This being said, the fitted value has the expected magnitude: using $\beta=1$ (thus making the value of $\tau_0$ irrelevant), $\alphas=0.3$ and $C_g=3$ for a fragmenting gluon with $\ave{z}_g=0.5$~\cite{deFlorian:2007aj} leads to $K_\textnormal{th} =0.43$. 

Eq.~\eqref{eq:eloss_scaling} can also be used to predict $\RAA$ in other collision systems, such as OO collisions at $\sqrt{s}=7\TeV$ planned at LHC Run 3~\cite{Brewer:2021kiv}. Using the nominal Glauber model and the multiplicity from \texttt{EPOS3.402}~\cite{Werner:2013tya} gives $\meanepsbar_{\,_{\textnormal{OO}}}=0.61 {\raisebox{0.5ex}{\tiny$\substack{+0.17\\-0.10}$}}$\GeV, leading to $R_{\textnormal{OO}}(\pt=20\GeV)=0.85{\raisebox{0.5ex}{\tiny$\substack{+0.04\\-0.02}$}}$ in minimum bias collisions.

Once the dependence of $\meanepsbar$ with $L$ is empirically determined, it becomes possible to investigate the azimuthal dependence of hadron suppression, from which the $v_2$ coefficient can be computed. Using \eqref{eq:RAA2} and \eqref{eq:eloss_scaling}, the $\phi$ dependence of \RAA\ can be modeled as
\be\label{eq:raa_phi}
\RAA(u, n, \phi) = f\left(u\times \left(L/L(\phi)\right)^\beta, n\right)\,,
\ee
where $L(\phi)$ is given by \eqref{eq:L} with $\bm{\ell}$ along $\phi$ and $\phi=0$ is the direction of the impact parameter $\bm{b}$. 
Let us assume that $L(\phi)$ can be approximated as
\be\label{eq:Lphi}
L(\phi) = L\times\left(1-\ecc\, \cos\left(2\phi\right)\right)\,,
\ee
where the eccentricity $\ecc$ is thus given by~\footnote{This definition of $\ecc$ is different than the usual eccentricity $\varepsilon=(\ave{y}^2-\ave{x}^2)/(\ave{y}^2+\ave{x}^2)$ used to scale $v_2$ at low $\pt$.}
\be\label{eq:ecc}
\ecc = \frac{L(\pi/2)-L(0)}{L(\pi/2)+L(0)}\,.
\ee
From the definition of the $v_m$ coefficients, we have
\bea
\frac{\RAA(u, n, \phi)}{\RAA(u, n)} &=& 1 + 2\, \sum_{m=1}^\infty v_{2m}\,\cos(2m\phi)\label{eq:def_vnall}\\
&\simeq& 1 + 2\, v_2\,\cos(2\phi)\label{eq:def_vn2}\,,
\eea
where in \eqref{eq:def_vn2} the higher order harmonics are neglected at high $\pt$~\cite{CMS:2017xgk}. From \eqref{eq:raa_phi}, \eqref{eq:Lphi} and \eqref{eq:def_vn2}, one gets
\bea\label{eq:vn}
2\,v_2 &\simeq& \frac{\RAA(0)-\RAA(\pi/2)}{\RAA(0)+\RAA(\pi/2)}\nonumber\\[0.2cm]
 &\simeq& \frac{f(u/(1-\ecc)^\beta)-f(u/(1+\ecc)^\beta)}{f(u/(1-\ecc)^\beta)+f(u/(1+\ecc)^\beta)}\,.
\eea
Performing the Taylor expansion of \eqref{eq:vn} to first order in $\ecc$ leads to
\bea
\frac{v_2(u, n)}{\ecc} &\simeq& \frac\beta{2}\,\frac{\partial \ln f(u, n)}{\partial \ln u}\,, \label{eq:v2finala}\\
\frac{v_2(\pt)}{\ecc}&\simeq& \frac{\beta}{2}\,\frac{\pt}{\RAA(\pt)}\,\frac{\partial \RAA(\pt)}{\partial \pt}\,.\label{eq:v2finalb}
\eea
Within the above assumptions, the quantity $v_2/\ecc$ at large $\pt$ is simply proportional to the logarithmic derivative of $\RAA$ and to the exponent $\beta$. As a consequence, $v_2/\ecc$ has the same universal dependence on $\pt/\meanepsbar$ as $\RAA$, for all collision energies and centrality classes. It is given by
\bea\label{eq:v2approx}
\frac{v_2(u, n)}{\ecc} &=& \frac{\beta}{2}\,\frac{n}{u}\,\int\,\dd x\, \bar{P}(x)\, \frac{x}{(1+x/u)^{n+1}}\Big/\nonumber \\&&\int\,\dd x\, \bar{P}(x)\, \frac{1}{(1+x/u)^n}\,,
\eea
using \eqref{eq:RAA2} in \eqref{eq:v2finala}. Eq.~\eqref{eq:v2finalb} moreover indicates that $v_2$ and $\RAA$ at a given $\pt$ are trivially related for measurements from the same collision system. In particular, this relation does not involve the knowledge of the energy loss scale $\meanepsbar$. Finally, the normalization uncertainties of $\RAA$ vanish when computing~\eqref{eq:v2finalb}.
\begin{figure}[t]
\begin{center}
    \includegraphics[width=0.47\textwidth]{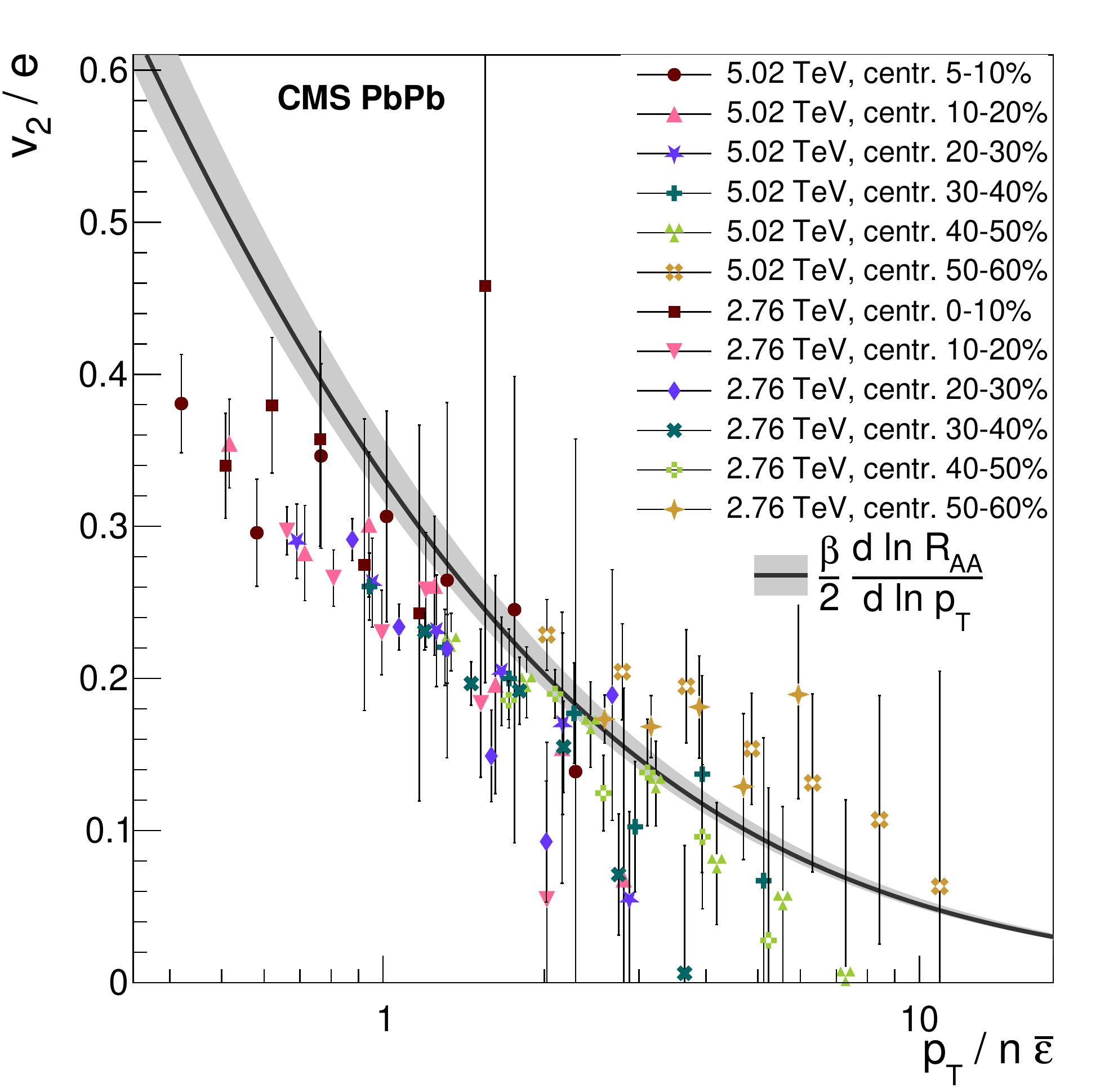}
  \end{center}
\vspace{-0.4cm}
\caption{Scaling of $v_2/\ecc$ of $h^\pm$ in \PbPb collisions at $\sqrt{s}=2.76$ and $5.02$\TeV~\cite{CMS:2012tqw,CMS:2017xgk} as a function of $\pt/n\meanepsbar$. The solid band $\beta=1.02^{+0.09}_{-0.06}$ shows Eq.~\eqref{eq:v2approx} using $n=5.5$.}
\label{fig:v2scaling}
\end{figure}

In each collision system, a full computation of $\RAA(\pt, \phi)$ has been performed starting from \eqref{eq:raa_phi} and using the nominal Glauber model, without any assumption on $L(\phi)$. Fitting $\RAA(\pt, \phi)$ with \eqref{eq:def_vnall}, restricted to the first three even harmonics, provides an `exact' coefficient $v_2$ within the model. We find that $v_2/\ecc$ (where $\ecc$ is computed in the Glauber model) is very well reproduced by the approximation \eqref{eq:v2approx}, plotted in Fig.~\ref{fig:v2scaling} (gray band), especially for centralities within $5-60\%$. It is maximal at low $\pt/n\meanepsbar$, smoothly decreases and vanishes in the large $\pt/n\meanepsbar$ limit when energy loss effects become negligible. The CMS measurements of $v_2$ in \PbPb collisions at $\sqrts=2.76$\TeV and $\sqrts=5.02$\TeV~\cite{CMS:2012tqw,CMS:2017xgk}, reaching up to $\pt\simeq 100\GeV$, are also shown in Fig.~\ref{fig:v2scaling} for $\pt>15\GeV$ as a function of the scaling variable $\pt/n\meanepsbar$, where the values of $\meanepsbar$ originate from the fits of $\RAA$. The predicted scaling for the different collision systems is clearly apparent~\footnote{The data in the $0$--$5\%$ centrality class, not shown here, are subject to important fluctuations~\cite{Miller:2003kd,Bhalerao:2006tp} that also affect the estimation of $\ecc$. They deviate from the scaling observed in the other classes.}. Note that the values of $\meanepsbar$ are fixed and no longer left as free parameters as in Fig.~\ref{fig:scaling}. The model reproduces well the data above $\pt/n\meanepsbar\gtrsim1$, while the stronger predicted slope at lower values can be understood from \eqref{eq:v2finalb}: the slope of $\RAA$ --~and therefore the value of $v_2/\ecc$~-- proves larger than in the data as path-length fluctuations are not included in the model (due to surface emission, $\RAA$ should be bounded from below). Perhaps surprisingly, the analytic model appears able to reproduce both $\RAA$ and $v_2$ measurements, at least above $\pt\gtrsim 15$\GeV. In other words, it does not face the so-called `$\RAA\otimes v_2$ puzzle' investigated by many groups over the last decade~\cite{PHENIX:2010nlr,Molnar:2013eqa,Xu:2014ica,Das:2015ana,Noronha-Hostler:2016eow,Andres:2019eus,Zigic:2019sth,Zhao:2021vmu,He:2022evt} and whose resolution might involve event-by-event fluctuations in the soft sector~\cite{Noronha-Hostler:2016eow}. The consistency between $\RAA$ and $v_2$ measurements within the present data-driven approach is thus compelling~--~moreover with a path-length dependence compatible with perturbative QCD, unlike the conclusions of Ref.~\cite{PHENIX:2010nlr} based on RHIC data.

\begin{figure}[tbp]
\begin{center}
    \includegraphics[width=0.47\textwidth]{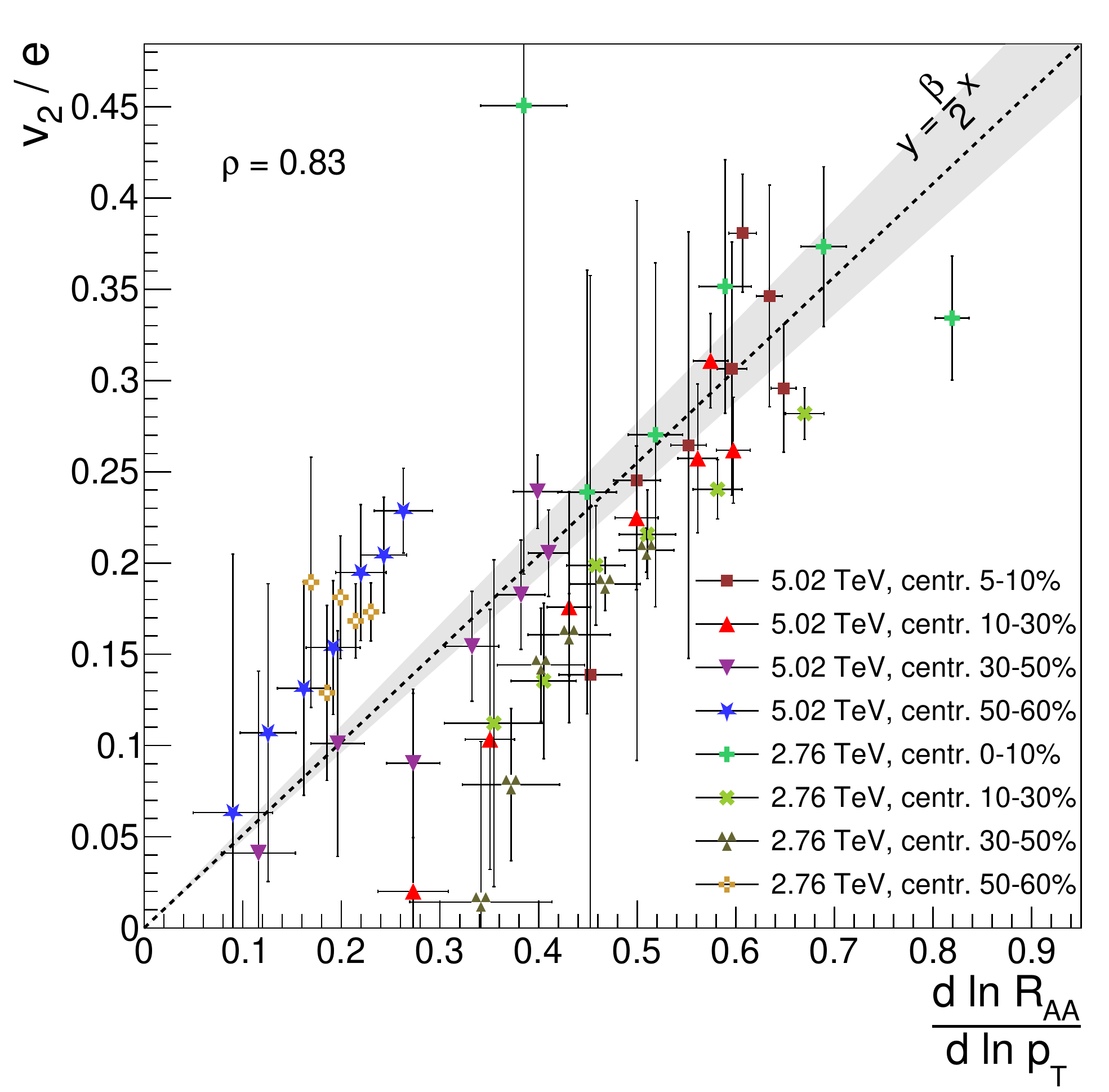}
  \end{center}
\vspace{-0.4cm}
\caption{Relation between $v_2/\ecc$~\cite{CMS:2012tqw,CMS:2017xgk} and $\dd\ln\RAA/\dd\ln\pt$~\cite{CMS:2012aa,CMS:2016xef} measured by CMS. The band \mbox{$y=\beta\,x/2$} (\mbox{$\beta=1.02^{+0.09}_{-0.06}$}) shows the expectation from~\eqref{eq:v2finalb}.}
\label{fig:v2vsRAA}
\end{figure}

While the agreement between the full calculation of $v_2/\ecc$ and Eq.~\eqref{eq:v2finalb} already suggests that this latter relation may be valid, we would like to check whether it also holds when comparing solely $\RAA$ and $v_2$ \emph{measurements}, independently of the present energy loss model. In order to reduce the uncertainty due to bin-to-bin statistical fluctuations, we perform an agnostic fit (using Chebyshev polynomials) of the CMS $\RAA$ data at $\sqrts=2.76$\TeV and~$5.02$\TeV~\cite{CMS:2012aa,CMS:2016xef}, from which the slope $\dd\ln\RAA/\dd\ln\pt$ is evaluated. The $v_2/\ecc$ measurements are plotted as a function of the $\RAA$ slope in Fig.~\ref{fig:v2vsRAA}. Although the present precision of the data do not allow yet for a rigorous test of Eq.~\eqref{eq:v2finalb}, the correlation between the two measured quantities is clearly apparent (correlation coefficient $\rho=0.83$). In addition, the expected function $y=\beta\,x/2$ (band in Fig.~\ref{fig:v2vsRAA}) reproduces fairly the observations, giving confidence that the relation between $\RAA$ and $v_2$  at large $\pt$ gives a direct experimental access to the path-length dependence of parton energy loss in QGP. A slight overshoot in the $50$--$60\%$ centrality class $v_2$ measurements may either signal back-to-back jet correlations contamination in data~\cite{CMS:2016xef} or the breakdown of~\eqref{eq:v2finalb}. In other centrality classes, disagreement may be due to the overestimation of \ecc in the Glauber model presently used, which relies on hard sphere nuclear densities.

In summary, the universal dependence of hadron $\RAA$ at high $\pt$ has been further checked using additional data sets from RHIC and LHC. Relating the values of $\meanepsbar$ extracted from $\RAA$ to the hadron multiplicity enables the determination of the $L$ dependence of parton energy loss, $\meanepsbar \propto L^{\beta}$ with $\beta=1.02{\raisebox{0.5ex}{\tiny$\substack{+0.09\\-0.06}$}}$, in agreement with a longitudinally expanding QGP. The $v_2/\ecc$ anisotropy coefficient exhibits the same scaling property as $\RAA$, in both the model and data. Finally the simple relation between $v_2/\ecc$ and $\RAA$ found in the model proved consistent with independent $v_2$ and $\RAA$ measurements, providing another access to the path-length dependence of parton energy loss. The LHC Run~3 should allow for testing with unprecedented precision these multiple scaling properties.

\begin{acknowledgments}
We thank Maxime Guilbaud for discussions. GF acknowledges the support from IN2P3
and William Miller, and the hospitality of Subatech where this work was completed. This work is funded by the ``Agence Nationale de la Recherche'' under grant ANR-18-CE31-0024-02.
\end{acknowledgments}

%

\end{document}